\documentclass[11pt,twoside]{article}
\usepackage{asp2010}
\usepackage{graphicx}

\resetcounters

\begin{document}

\title{Photometric properties of carbon stars in the Small Magellanic Cloud}
\author{G.~C.\ Sloan$^1$, E.\ Lagadec$^2$, K.~E.\ Kraemer$^3$, M.~L.\ 
Boyer$^4$, S.\ Srinivasan$^5$, I.\ McDonald$^6$, and A.~A.\ Zijlstra$^6$
\affil{$^1$Center for Radiophysics and Space Sciences, Cornell University,
  Ithaca, NY 14853-6081, USA}
\affil{$^2$Laboratoire Lagrange, UMR 7293, Univ.\ Nice Sophia-Antipolis, 
  CNRS, Observatoire de la C\^{o}te d'Azur, 06300, Nice, France}
\affil{$^3$Institute for Scientific Research, Boston College, Chestnut
  Hill, MA 02467, USA}
\affil{$^4$Observational Cosmology Lab, Code 665, NASA Goddard Space Flight 
  Center, Greenbelt, MD 20771, USA}
\affil{$^5$Academia Sinica Institute of Astronomy and Astrophysics, 
  Taiwan University, Roosevelt Road, Taipei 10617, Taiwan}
\affil{$^6$Jodrell Bank Centre for Astrophysics, Alan Turing Building, 
  Manchester, M13 9PL, UK} }

\begin{abstract}

The Optical Gravitational Lensing Experiment identified over 
1,800 carbon-rich Mira and semi-regular variables in the 
Small Magellanic Cloud.  Multi-epoch infrared photometry 
reveals that the semi-regulars and Miras follow different 
sequences in color-color space when using colors sensitive to 
molecular absorption bands.  The dustiest Miras have the 
strongest pulsation amplitudes and longest periods.  Efforts 
to determine bolometric magnitudes reveal possible systematic 
errors with published bolometric corrections.

\end{abstract}

%\section{Introduction}

The Optical Gravitional Lensing Experiment (OGLE) has 
surveyed the Small Magellanic Cloud (SMC) for variables and 
transients.  The OGLE-III experiment discovered over 4,500 
Mira and semi-regular variables (SRVs).  Carbon stars account 
for 315 of the Miras and 1,488 of the semi-regular variables 
(Soszy\'{n}ski et al. 2011).

To investigate the infrared (IR) photometric properties of 
the carbon-rich long-period variables (LPVs), we have 
searched multiple archival databases to generate a 
time-averaged spectral energy distribution for each source.
The OGLE survey provides mean $V$ and $I$ data.  
The Two-Micron All-Sky Survey (2MASS), the deeper 2MASS 6x 
survey, and the Deep Near-Infrared Survey of the Southern Sky 
(DENIS) provide three or more epochs at $J$ and $K$ and two 
at $H$ (Skrutskie et al.\ 2006; Cioni et al.\ 2000).  The 
SAGE-SMC survey (Surveying the Agents of Galactic Evolution), 
in combination with the S$^3$MC survey ({\it Spitzer} Survey 
of the SMC), provides three epochs at 3.6, 4.5, 5.8, 8.0, and 
24~$\mu$m in the core of the SMC and two throughout the 
galaxy (Gordon et al.\ 2011; Bolatto et al.\ 2007).  We have 
also turned to the Wide-field Infrared Survey Experiment 
(WISE) for additional temporal coverage at 3.4 and 4.6~$\mu$m 
(Wright et al.\ 2010), which we use in conjunction with the 
{\it Spitzer} data at 3.6 and 4.5~$\mu$m.

Our presentation in Vienna considered both the Large and
Small Magellanic Clouds, but here, due to space restrictions,
we focus on just the SMC.  The results for the two galaxies 
are generally very similar.

\begin{figure}[!ht] % Fig. 1
\centerline{
\includegraphics[width=250pt]{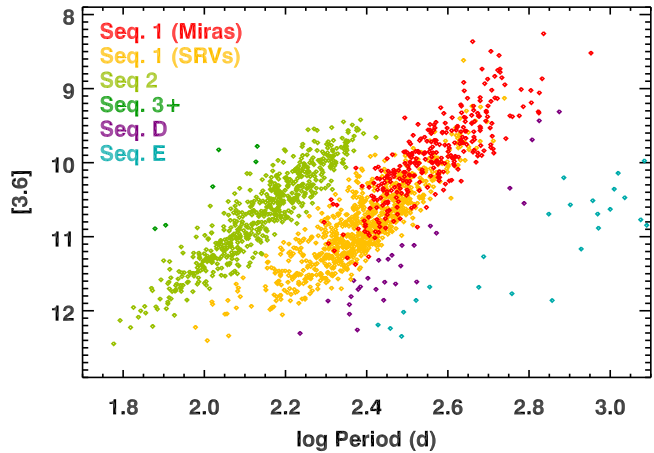}
}
\caption{\label{fig:1} The period-luminosity diagram for 
carbon-rich Miras and SRVs, color-coded by pulsational 
sequence (following the nomenclature of Fraser et al.\ 2008).
Fundamental-mode pulsators on the Sequence 1 are coded red 
and orange depending on whether the OGLE survey identified 
them as Miras or SRVs.}
\end{figure}

\begin{figure}[!ht] % Fig. 2
\centerline{
\includegraphics[width=250pt]{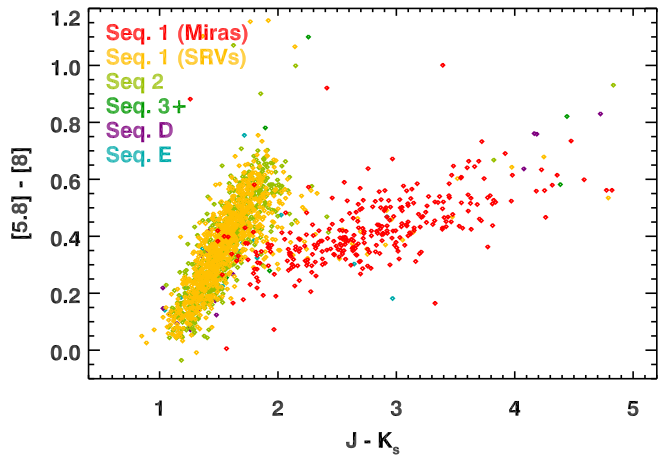}
}
\caption{\label{fig:2} In the [5.8]$-$[8] vs.\ $J$$-$$K_s$ 
plane, nearly all semi-regulars follow a blue sequence with 
$J$$-$$K_s$ $\la$ 2, while Miras dominate the redder 
sequence.}
\end{figure}

The OGLE-III survey provides three periods and amplitudes.
We chose the first period and amplitude corresponding to 
Sequence 1-4 (adopting the nomenclature of Fraser et al.\ 
2008).  Figure 1 shows how most of the Miras and SRVs fall 
along Sequences 1 and 2, which are the fundamental pulsation 
mode and the first overtone, respectively (Wood \& Sebo 
1996).  These sequences are the basis for the color-coding in 
Figures 1 and 2, with different colors to distinguish the 
Miras and SRVs on Sequence 1.  The OGLE-III survey separates 
these two variability classes at an $I$-band amplitude of 0.8 
mag (peak-to-peak).  Figure 2 shows how [5.8]$-$[8] and 
$J$$-$$K_s$ colors separate the Miras and SRVs relatively 
cleanly.  Most of the SRVs fall on a sequence where 
[5.8]$-$[8] increases quickly as $J$$-$$K_s$ increases, while 
most of the Miras fall on a sequence which is much redder in 
$J$$-$$K_s$ and has a much shallower slope.  

\begin{figure} % Fig. 3
\centerline{
\includegraphics[width=240pt]{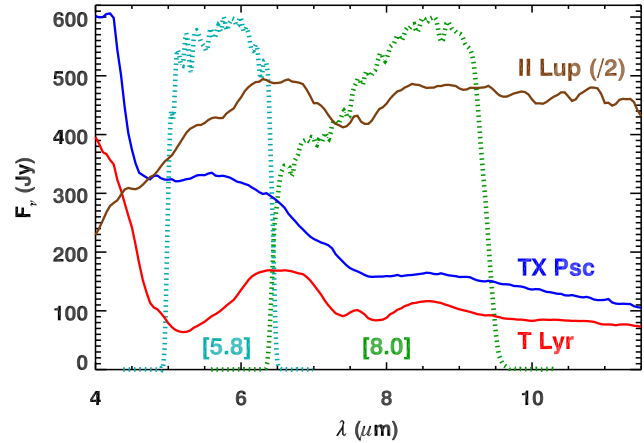}
}
\caption{\label{fig:3} The 5.8 and 8~$\mu$m bandpasses on 
the Infrared Array Camera (IRAC) on {\it Spitzer} plotted
over spectra from the spectral atlas from the Short-Wavelength 
Spectrometer (SWS) on the {\it Infrared Space Observatory}
(Sloan et al.\ 2003).  T Lyr has a redder [5.8]$-$[8]
color than TX Psc because of the deep absorption from C$_3$
at $\sim$5~$\mu$m.  II Lup is red because of its thick dust
shell.}
\end{figure}

Figure 3 shows how the infrared spectra lead to two 
sequences, which we will refer to as the SRV and Mira 
sequences.  Redder [5.8]$-$[8] colors on the SRV sequence,
represented by TX Psc and T Lyr,
result from increasing absorption from C$_3$ at 
4.5--6.0~$\mu$m, which affects the 5.8~$\mu$m bandpass.  
On the Mira sequence, represented by II Lup, redder colors 
result from increasing dust opacity.

\begin{figure}[!ht] % Fig. 4
\centerline{
\includegraphics[width=248pt]{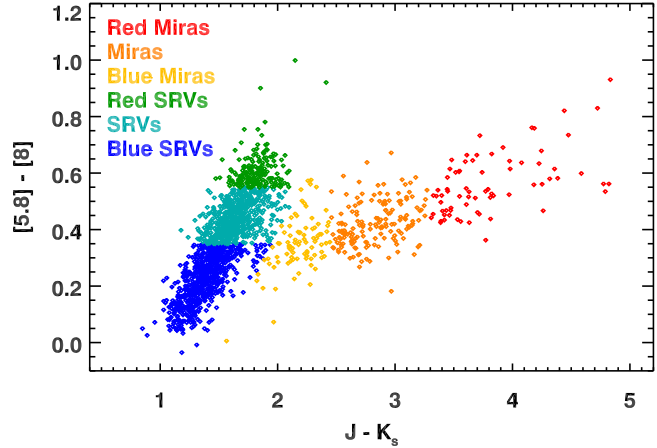}
}
\caption{\label{fig:4} Carbon-rich Miras and SRVs in the SMC, 
color-coded by position along the SRV and Mira sequences in 
[5.8]$-$[8] vs.\ $J$$-$$K_s$ space.  Both sequences are 
divided into three strips to define the six groups.}
\end{figure}

\begin{figure}[!ht] % Fig. 5
\centerline{
\includegraphics[width=248pt]{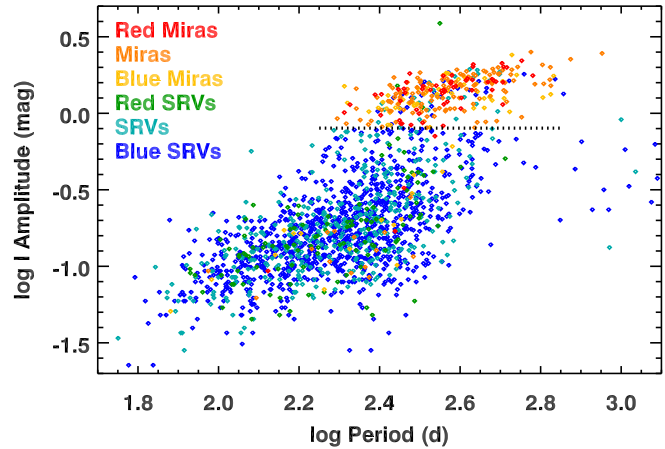}
}
\caption{\label{fig:5} Pulsational amplitude at $I$ vs.\ 
period, color-coded as in Fig.\ 3.  The sources on the Mira 
sequence have larger amplitudes, and for these amplitudes, 
the reddest Miras have the longest periods.  The horizontal 
dashed line shows the amplitude boundary of 0.80 mag between
Miras and SRVs in the OGLE-III survey.}
\end{figure}

For the remaining figures in this contribution, we have 
adopted the color-coding defined in Figure 4, with blue to 
green depicting the SRV sequence in order of increasing 
[5.8]$-$[8] color and yellow to red tracking increasing 
$J$$-$$K_s$ color along the Mira sequence.  Figure 5 shows 
how these different groups map into pulsation amplitude at 
$I$ and pulsation period.  These groups improve on the 
boundary between Miras and SRVs adopted by the OGLE-III 
survey (at $\Delta$$I$ = 0.8 mag).  The Mira sequence is 
associated with the larger amplitudes, and within this group, 
the dustier sources have the longest periods.  This figure 
clearly links pulsation and dust production.

\begin{figure}[!ht] % Fig. 6
\centerline{
\includegraphics[width=248pt]{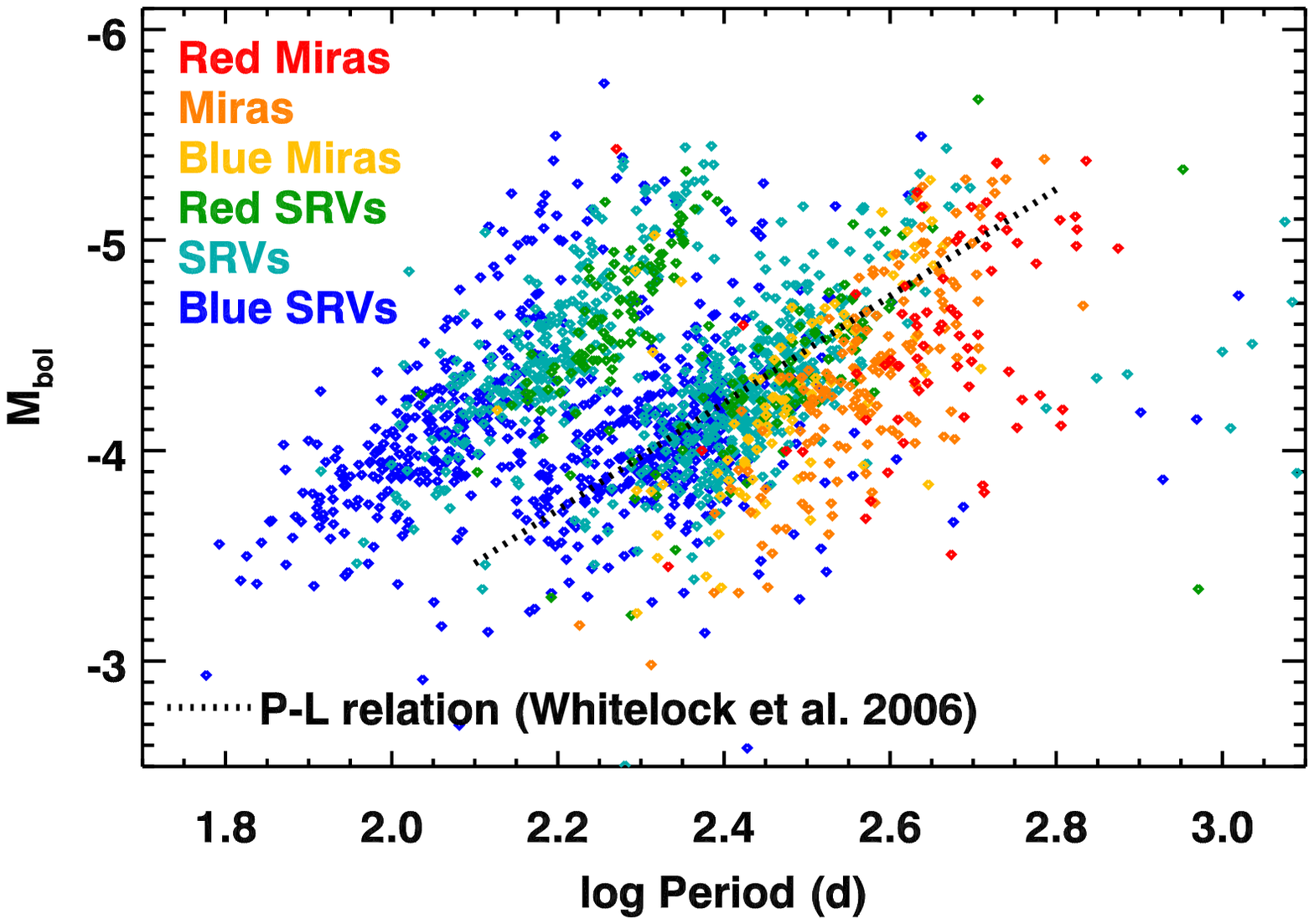}
}
\caption{\label{fig:6} The period-luminosity diagram for the
SMC after making $J$$-$$K_s$-based bolometric corrections,
color-coded as defined in Fig.\ 4.  With these BCs, sources
from different parts of the Mira and SRV sequences shift away 
from each other in P-L space, creating the {\it blue slip} 
and {\it red droop}.}
\end{figure}

\begin{figure}[!ht] % Fig. 7
\centerline{
\includegraphics[width=248pt]{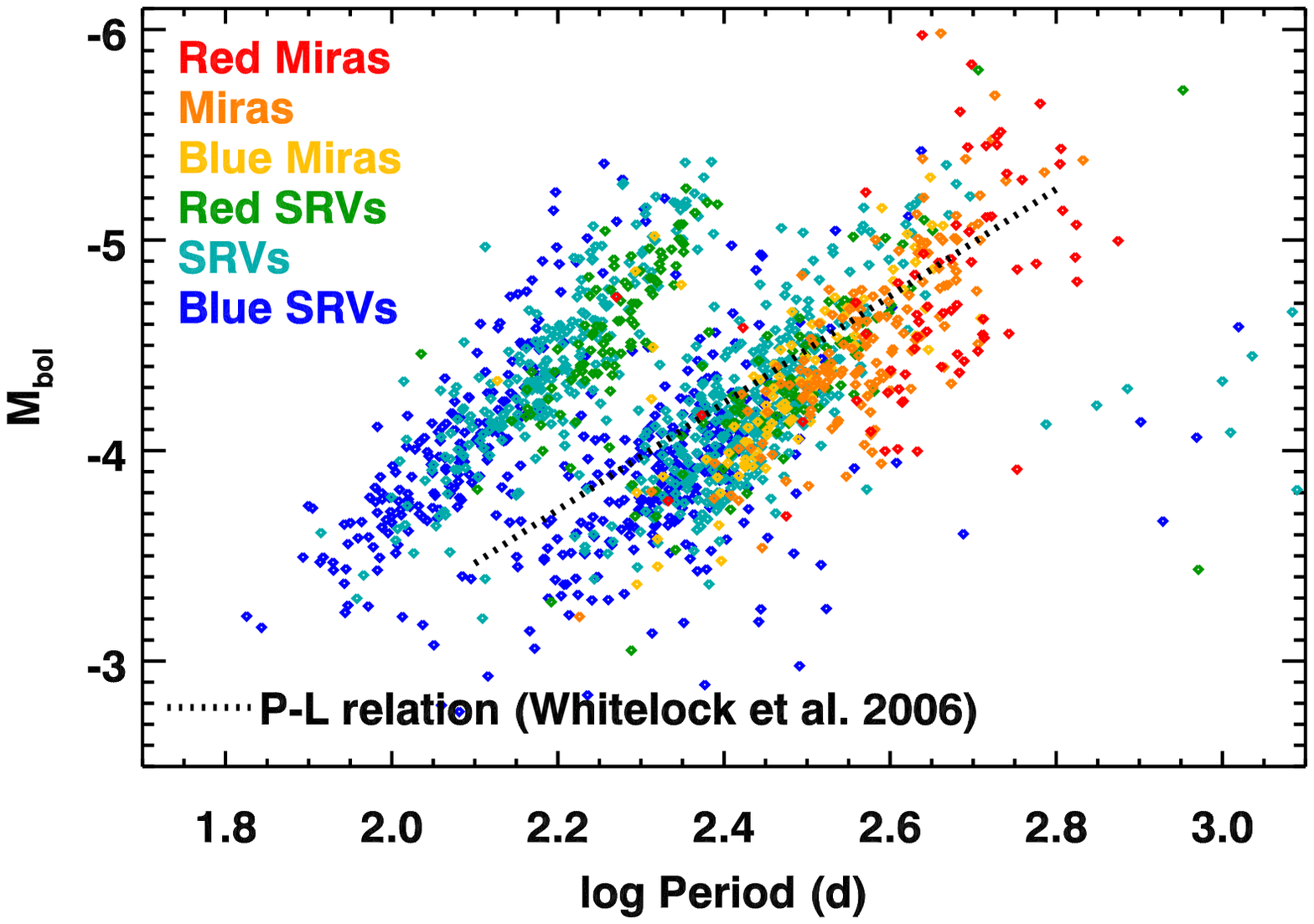}
}
\caption{\label{fig:7} Bolometric corrections based on 
$K$$-$$L$ reduce both the {\it blue slip} and the {\it red 
droop}, but they do not eliminate them completely.}
\end{figure}

Comparing the bolometric magnitude and pulsation to 
theoretical evolutionary tracks allows one to estimate 
initial masses, but determining the bolometric magnitude
from the available photometry is non-trivial.  In Figure 6,
the bolometric magnitudes are determined with bolometric
corrections (BCs) based on $J$$-$$K_s$ (Whitelock et al.\
2006).  Two problems are apparent.  The reddest and bluest 
sources on the SRV sequence have separated from each other, 
on both the overtone and fundamental modes.  We call this 
problem the {\it blue slip}.  It should be noted that we have 
applied the BC below the recommended limit of $J$$-$$K_s$ 
$\ga$ 1.5.  On the fundamental mode, another problem which we 
describe as the {\it red droop} has shifted the apparent 
bolometric magnitudes of the reddest sources well below the 
expected P-L relation.  Using BCs from other authors produces 
results similar to those illustrated here.

Figure 7 shows that BCs based on $K$$-$$L$ reduce both the 
{\it blue slip} and {\it red droop}, but not entirely in 
either case.  (We have treated $L$, [3.4], and [3.6] as
equivalent here.)  These problems probably arise from the BCs 
and are not intrinsic to the stars themselves.  The need for 
improved means of determining bolometric magnitudes is readily 
apparent.

%\acknowledgements Thank you world!

\vspace{0.5cm}

\bibliography{aspauthor}

\noindent Bolatto, A.~D., Simon, J.~D., Stanimirovi\'{c}, S., et al.\ 2007, 
  ApJ, 655, 212\\
\noindent Cioni, M.-R., Loup, C., Habing, H.~J., et al.\ 2000, A\&AS, 144, 235\\
\noindent Fraser, O.~J., Hawley, S.~L., \& Cook, K.~H.\ 2008, AJ, 136, 1242\\
\noindent Gordon, K.~D., Meixner, M., Meade, M.~R., et al.\ 2011, AJ, 142, 102\\
\noindent Skrutskie, M.~F., Cutri, R.~M., Stiening, R., et al.\ 2006, AJ, 131, 
  1163\\
\noindent Sloan, G.~C., Kraemer, K.~E., Price, S.~D., \& Shipman, R.~F.\ 2003, 
  ApJS, 147, 379\\
\noindent Soszy\'{n}ski, I., Udalski, A., Szyma\'{n}ski, M.~K., et al.\ 2011,
  Act.\ Astron.\ 61, 217\\
\noindent Whitelock, P.~A., Feast, M.~W., Marang, F., \& Groenewegen, M.~A.~T.\
  2006, MNRAS,\\
\indent 369, 751\\
\noindent Wood, P.~R., \& Sebo, K.~M.\ 1996, MNRAS, 282, 958\\
\noindent Wright, E.~L., Eisenhardt, P.~R.~M., Mainzer, A.~K., et al.\ 2010, 
  AJ, 140, 1868

\end{document}